\documentclass[manuscript]{aastex}
\usepackage{emulateapj5}

\shorttitle{SNR G57.2+0.8 and SGR\,1935+2154}
\shortauthors{Kothes et al.}

\begin{document}

\title{Radio continuum and polarization study of SNR G57.2+0.8 associated with magnetar 
SGR\,1935+2154}

\author{R. Kothes\altaffilmark{1}, X. Sun\altaffilmark{2}, 
B. Gaensler\altaffilmark{2,3},
\& W. Reich\altaffilmark{4}}

\altaffiltext{1}{National Research Council Canada,
              Herzberg Programs in Astronomy \& Astrophysics,
              Dominion Radio Astrophysical Observatory,
              P.O. Box 248, Penticton, British Columbia, V2A 6J9, Canada}

\altaffiltext{2}{Department of Astronomy, Yunnan University, and Key Laboratory of Astroparticle Physics of Yunnan Province, Kunming, 650091, China}
%\altaffiltext{2}{The University of Sydney, NSW 2006, Australia}

%\altaffiltext{3}{Department of Physics and Astronomy, Brandon University, 
%             270 18th Street, Brandon, MB R7A 6A9, Canada}
   
\altaffiltext{3}{Dunlap Institute for Astronomy and Astrophysics,
            University of Toronto, 50 St. George Street, Toronto, ON M5S 3H4, Canada}
   
\altaffiltext{4}{Max-Planck-Institut f\"ur Radioastronomie, 
            Auf dem H\"ugel 69, D-53121
            Bonn, Germany}

\email{roland.kothes@nrc-cnrc.gc.ca}

\begin{abstract}
We present a radio continuum and linear polarization study of the Galactic supernova remnant
G57.2+0.8, which may host the recently discovered magnetar SGR\,1935+2154. The radio SNR shows
the typical radio continuum spectrum of a mature supernova remnant with a spectral index
of $\alpha = -0.55 \pm 0.02$ and moderate polarized intensity. Magnetic field vectors indicate a
tangential magnetic field, expected for an evolved SNR, in one part of the SNR and a 
radial magnetic field in the other. The latter can be explained by an overlapping 
arc-like feature, perhaps a pulsar wind nebula, emanating from the magnetar. The presence of 
a pulsar wind nebula is supported by the low average braking index of 1.2, we extrapolated for the magnetar,
and the detection of diffuse X-ray emission around it.
We found a distance of 12.5~kpc for the SNR,
which identifies G57.2+0.8 as a resident of the Outer spiral arm of the Milky Way. The SNR has a radius of
about 20~pc and could be as old as 41,000 years. The SNR has already entered the radiative 
or pressure-driven snowplow phase of
its evolution. We compared independently determined characteristics like age and distance for both, the SNR
and SGR\,1935+2154, and conclude that they are physically related.
%We propose a B-type progenitor star,
%which did not produce a significant stellar wind during its lifetime similar to SNR CTB\,109 and 
%its magnetar. 
\end{abstract}


\section{Introduction}

Magnetars are young neutron stars with extreme surface magnetic fields, characterised by intense episodes 
of X-ray and $\gamma$-ray flaring \citep{wt06,kb17}. They are windows into extreme physical processes, and 
may also be the progenitor of time-domain phenomena such as long gamma-ray bursts and fast radio bursts 
\citep[e.g.,][]{tcq04,kat16,mbm17}. However, the number of known magnetars is still relatively small 
\citep{ok14}, and their formation mechanisms, birth rate, evolution, and relation to other types of 
neutron stars are all still not understood.

Key information on magnetars can be obtained by studying their environments. Specifically, several magnetars 
are associated with supernova remnants (SNRs) \citep{gsgv01}, which provides the opportunity to make 
independent age and distance estimates. Each new magnetar/SNR association helps characterize the 
energetics and lifetime of magnetar outbursts, and maps out the progenitor and supernova phase space that 
can subsequently produce a magnetar.

Soft Gamma Repeater (SGR)~1935+2154 (originally named GRB~140705A) was discovered on July 14, 2005, when 
the {\em Swift}\ satellite detected a burst of gamma-rays and an accompanying previously unidentified X-ray 
source \citep{smps14}. The source's low Galactic latitude and soft gamma-ray spectrum led \cite{lbb+14} 
to identify the source as a likely new magnetar, an identification that was confirmed when \cite{irz+14} 
detected a 3.2-second periodicity in the X-ray emission. 

\cite{gaen14} compared the position of SGR~1935+2154 to the SNR catalog of \cite{gre14}, and pointed out 
that SGR~1935+2154 sits close to the geometric center of the Galactic SNR G57.2+0.8.

G57.2+0.8 was first identified by \citet{sieb84} in a radio continuum survey of the area surrounding 
the (unrelated) millisecond pulsar B1937+21, and seems to be a reasonably typical shell-type SNR. 
\cite{gaen14} noted that there were few SNRs in this part of the Galactic plane, suggesting a low 
probability of a chance coincidence between the magnetar and the SNR, and thus a good likelihood of 
a physical association. However, few properties of the SNR have been established, and given the 
history of spurious magnetar/SNR associations, the possibility of a claimed association needs 
to be carefully assessed. 

Here we present new and archival radio data on this SNR, covering radio continuum, radio polarization 
and 21~cm HI observations. In Section~2, we we describe the observations and data processing procedures 
of the Effelsberg and DRAO synthesis telescope data and present a description of the results in Section~3.
An analysis of the results and their interpretation, including a thorough discussion on the possible 
association between SNR G57.2+0.8 and SGR\,1935+2154 is discussed in Section 4. A summary and conclusions
are provided in Section~5.

\section{Observations and Data Processing}

\subsection{Observations with the DRAO Synthesis Telescope}

1420~MHz and 408~MHz observations were obtained with the Dominion Radio Astrophysical 
Observatory's Synthesis
Telescope \citep[DRAO ST,][]{land00} as part of the Canadian Galactic Plane Survey 
\citep[CGPS,][]{tayl03}. Individual fields were processed using the 
routines described by \citet{will99} before they were mosaiced to the final
data products. To assure accurate representation to the largest scales
data observed with single antenna telescopes were incorporated after suitable 
filtering in the Fourier domain.
Continuum single antenna data were derived from the 408~MHz all-sky survey of
\cite{hasl82}, from the 1.4 GHz Effelsberg survey \citep{reic90}, and from
the second part of the Low-Resolution DRAO Survey of HI Emission from the 
Galactic Plane \citep{higg05} observed with the 26m John A. Galt telescope at DRAO. 

For the linear polarization data, no single antenna data were added. The
DRAO ST is sensitive to all structures from the resolution limit
(${\sim}1'$) up to $45'$. The absence of single-antenna data is not a
concern for observations of G57.2+0.8, whose maximum extent is smaller
than $15\arcmin$.
The DRAO Synthesis Telescope provides observations of linearly polarized emission at four frequency bands
around the HI line at 1420~MHz to allow precise determination of rotation
measures. Those bands are 7.5~MHz wide and the central frequencies are 1406.9~MHz for
band A, 1413.8~MHz for band B, 1427.4 MHz for band C, and 1434.3 MHz for band D.

The angular resolution of the CGPS data varies slightly across the final maps as
cosec($\delta$). At the centre of G57.2+0.8 we find a resolution of 
$8\farcm5 \times 2\farcm8$ for 408~MHz radio continuum,
$2\farcm5 \times 0\farcm82$ for 1420~MHz radio continuum, and $2\farcm9 \times 0\farcm97$
for the HI data.

\subsection{Observations with the Effelsberg Telescope}

We conducted radio continuum observations of G57.2+0.8 at high radio frequencies including
linear polarization with the Effelsberg 100-m radio telescope 
centered at $\alpha_{1950} = 19^h 32^m 50^s$ and $\delta_{1950} = 21\degr 50^{\prime}$.
We observed the SNR in July 1996 with the two feed 4.85~GHz receiver. In September of the same
year the four feed 10.45~GHz receiver was used for three coverages of G57.2+0.8 at
different parallactic angles which were
combined later. At both frequencies circluarly polarized components were recorded to
obtain total intensity and, by IF correlation, Stokes U and Q parameters. 

All observations
were made in the equatorial coordinate system. The scan direction was in azimuth.
The standard data 
reduction software
package based on the NOD2 format has been applied \citep{hasl74}. Individual multi-feed
observations were restored by averaging the coverages obtained with the different feed.
Baseline improvements by unsharp
masking \citep{sofu79} were applied to the observations. The ``Plait''
algorithm described by \citet{emer88} was
used to combine coverages observed at different
parallactic angles by destriping the maps in the Fourier domain. This
increases the signal-to-noise ratio of the final map substantially.

The 8.35~GHz (3.6~cm) single-beam Effelsberg observations were done
in May/June 2004 for total and polarized intensities. The G57.2 field
was extracted from three sets of long scans running perpendicular
across the Galactic plane. For details of the 8.35 GHz receiver see 
\citet{koth06b}. The resulting total intensity and polarization
maps are shown in Figs.~\ref{fig:tp} and \ref{fig:pol}. 

G57.2+0.8 is included in the Effelsberg 11-cm (2.7~GHz) Galactic
plane survey \citep{reic84}. It was again observed at 2.64~GHz
during test observations of a new 11~cm receiver with 80~MHz
bandwidth and a higher sensitivity, including linear polarization,
in June 2006. The receiver concept follows those of the 4.85 GHz
and the 10.45 GHz receivers. A small map centred on G57.2+0.8 was
composed from 10$\degr$ long scans crossing the Galactic plane. At a
beamwidth of 4.4$\arcmin$ G57.2+0.8 remains unresolved, thus providing
total flux density, percentage polarization and polarization angle.

The resolution in the final maps is $2\farcm6$ at 4.85~GHz and $1\farcm4$ at 8.35~GHz. The 
10.45~GHz observations 
were convolved to $1\farcm5$ to increase the signal-to-noise ratio. We found an rms sensitivity 
of 0.8~mJy\,beam$^{-1}$ in total power and 0.4~mJy\,beam$^{-1}$ in polarization at 10.45~GHz,
about 1.3~mJy\,beam$^{-1}$ in total power and 0.7~mJy\,beam$^{-1}$ in polarization at 8.35~GHz,
and 1.5~mJy\,beam$^{-1}$ in total power and 0.5~mJy\,beam$^{-1}$ in polarization at 4.85~GHz.
At 8.35~GHz the given rms sensitivity is an average, since there is a rms gradient from East
(low noise) to West (High noise), with a change of about a factor of 2 in sensitivity.

\section{Results}

\subsection{Radio Continuum Emission}

\begin{figure}[!ht]
\centerline{
  \includegraphics[bb = 40 60 480 705,height=19cm,angle=-90]{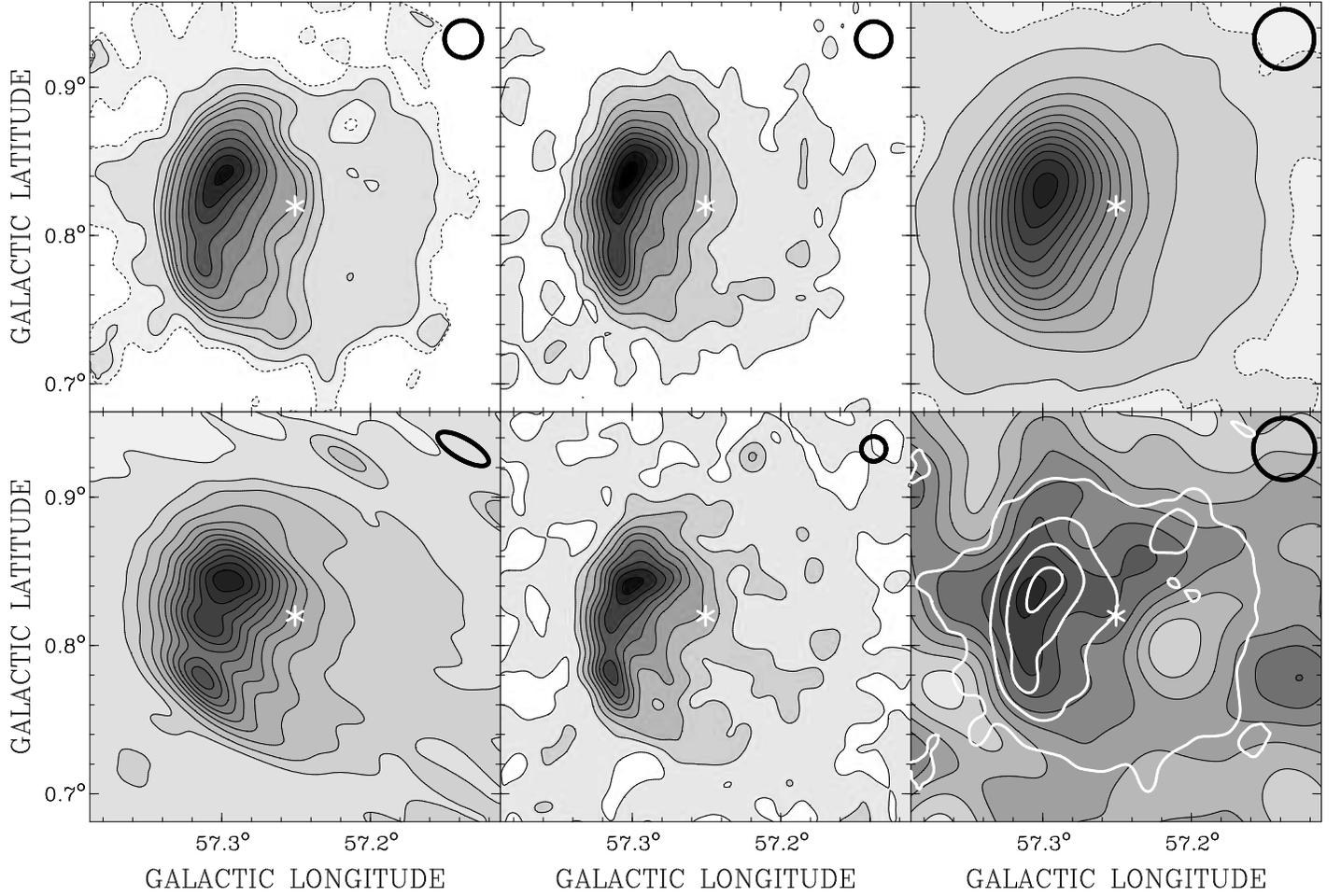}}
\caption{\label{fig:tp} Total power maps of the SNR G57.2+0.8 at 10.45~GHz (top left), 8.35~GHz (top centre),
4.85~GHz (top right), 1420~MHz CGPS (bottom left), 1420~MHz VGPS (bottom centre), and 74~MHz (bottom right). 
%Black solid contours are at 3, 7,
%11, 15, 20, 30, 40, 50, 60, 68, and 75~K at 10.45~GHz; at 10, 30, 50, 70, 90, 110, 130, 150,
%170, 190, 210, and 230~K at 4.85~GHz; at 7.5, 7.8, 8.2, 9, 10, 11, 12, 13, 14, 15, 16, 17, and 
%18~K; and at 
The white asterisk indicates the location of the newly discovered magnetar SGR\,1935+2154. The white
contours in the 74~MHz image represent the structure in total intensity at 10.45~GHz in the top left panel. The
resolution of the maps is indicated in the top right corner of each panel.}
\end{figure}

Total power images of G57.2+0.8 are shown in Figure~\ref{fig:tp}. In
addition to our Effelsberg observations and the CGPS 1420~MHz data
we also show the SNR at 1420~MHz taken from the VLA Galactic Plane Survey
\citep[VGPS,][]{stil06} and at 74~MHz, taken from the VLA Low-Frequency 
Sky Survey Redux \citep[VLSSr,][]{vlss}. The
SNR is an almost circular source with an average diameter of about $10\arcmin$. 
It consists of one prominent shell to the east ontop of a smooth emission plateau
with little sub-structure. The geometric centre is at $\ell = 57\fdg24$ 
and $b = 0\fdg81$, very close to the recently discovered magnetar SGR\,1935+2154 at
$\ell = 57\fdg25$ and $b = 0\fdg82$. 

\begin{figure}[!ht]
\centerline{
 \includegraphics[bb = 50 50 455 455,width=10cm,clip=]{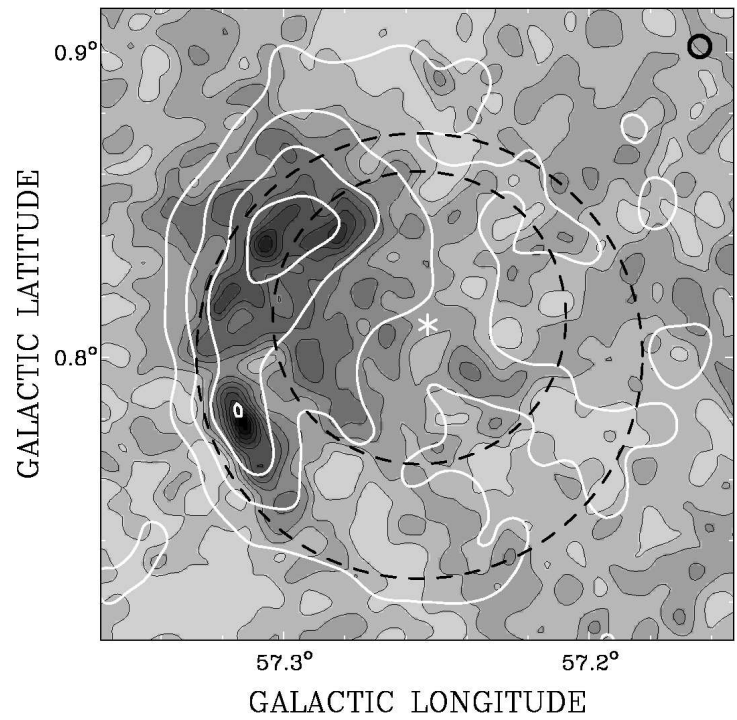}}
\caption{\label{fig:tgss} High resolution radio image taken from the first full data release of 
a survey of the 150~MHz radio sky, observed with the Giant Metrewave Radio Telescope (GMRT) as part of the 
TIFR GMRT Sky Survey (TGSS) project \citep{inte17}. White contours represent the VGPS observation 
(see Fig.~\ref{fig:tp}). The white asterisk marks the location of the magnetar SGR\,1935+2154. The
resolution of the map is 25" as indicated in the top right corner of the image. The two dashed circles
indicate the locations of the two arc-like features (see text).}
\end{figure}

In Fig.~\ref{fig:tgss} we display a radio map taken at 150~MHz with the GMRT as part of
the TGSS \citep{inte17}. This is the radio map with the highest angular resolution, however, it is not
sensitive to structures larger than a few arcminutes. Therefore we cannot extract an
integrated flux density from this survey, but can study the detailed structures of the 
bright shell. The shell is divided into two parts with a clear gap in between. The southern 
part looks like a highly compressed shell, typical for a mature shell-type SNR. The northern 
part is more complex. It might consist of two separate arc-like features. The curvature
of the inner one seems to be centered at the magnetar SGR\,1935+2154 and the outer arc 
could be a continuation of the southern shell. There is a lot of underlaying structure in
the Northern part that may be the result of these two features interacting.

\begin{figure}[!ht]
\centerline{
  \includegraphics[bb = 78 180 530 630,width=15cm,clip=]{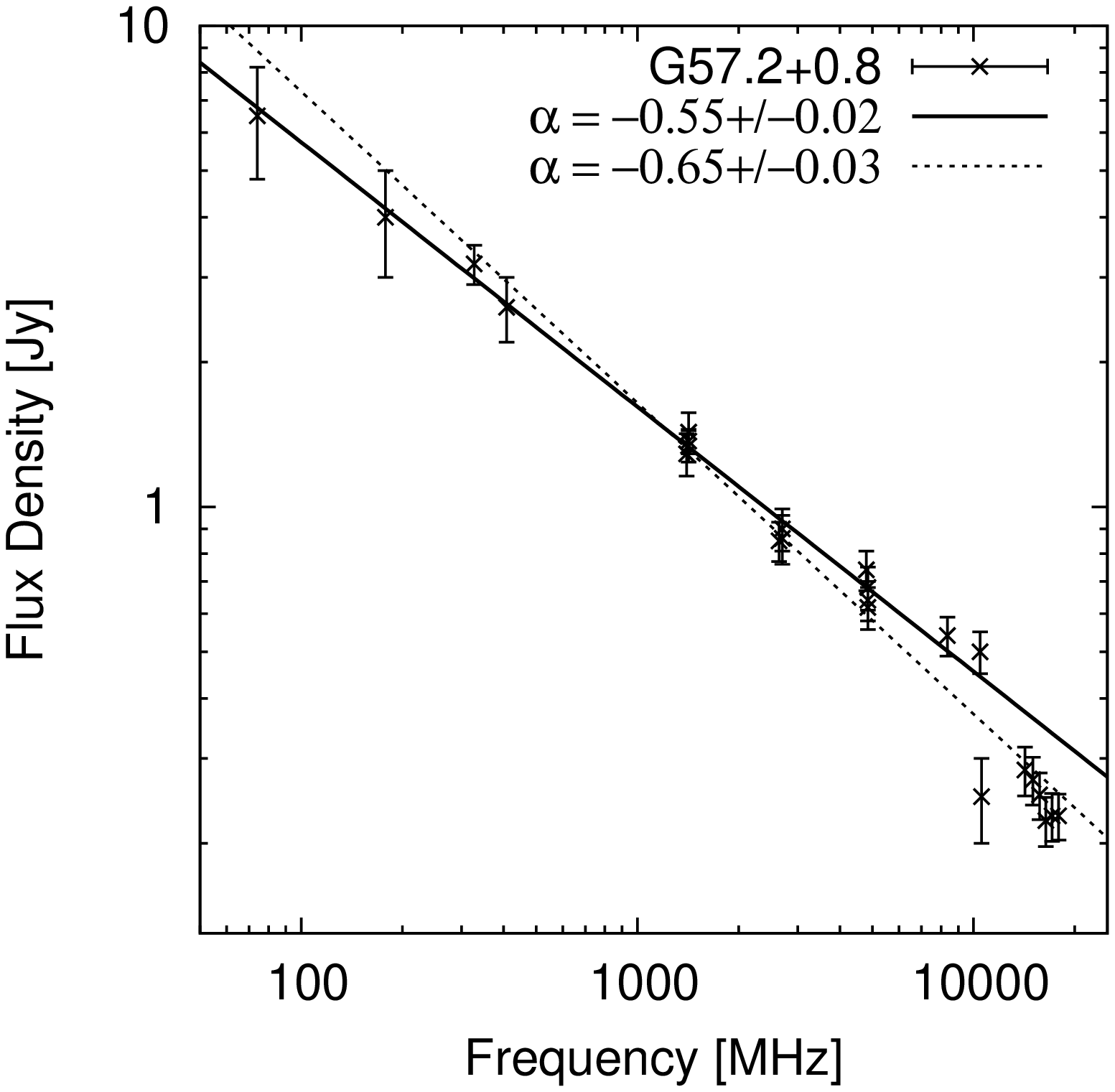}}
\caption{\label{fig:spec} Radio continuum spectrum of G57.2+9.8. Newly determined flux densities
are listed in Table~\ref{tab:fluxes}. Archival data were taken from 
\citet{reic84,sieb84,casw85,reic90,beck91,tayl92,whit92,hurl09,sun11,reic14}.}
\end{figure}

We determined flux densities from our observations and the archival 74~MHz VLSSr 
data \citep{vlss}. The flux densities were integrated 
in concentric rings centered at the location of the magnetar SGR\,1935+2154.
Newly determined flux densities are listed in Table~\ref{tab:fluxes}. This
table also contains new flux densities determined from an Effelsberg map at 2639~MHz and the 408~MHz CGPS data set,
which are not displayed in Figure~\ref{fig:tp}.
The radio continuum spectrum of G57.2+0.8 is displayed in Figure~\ref{fig:spec}.
The radio spectral index $\alpha$ is either $-0.55\pm 0.02$ or $-0.65\pm 0.03$, depending on 
whether we include
the AMI high frequency measurements \citep{hurl09} and the Effelsberg 10.7~GHz observation
by \citet{sieb84}. The \citet{sieb84} 10.7~GHz image does not cover the entire SNR,
only the bright shell. Therefore this can only be considered a lower limit. 
\citet{hurl09} explain the apparent lower flux values in the AMI
measurments by missing short spacings. It may also be that the
full extent of the source was underestimated, because only the shell part is an obvious
feature in the old observations. This
would lead to an overestimate for the subtracted background emission. Therefore we take
$\alpha= -0.55\pm 0.02$ as the better estimate of the true spectral index of G57.2+0.8 over
the frequency range displayed in Fig.~\ref{fig:spec}.
 
\subsection{Polarization and Rotation Measure}

\begin{figure}[!ht]
\centerline{
  \includegraphics[bb = 65 120 510 560,width=16cm]{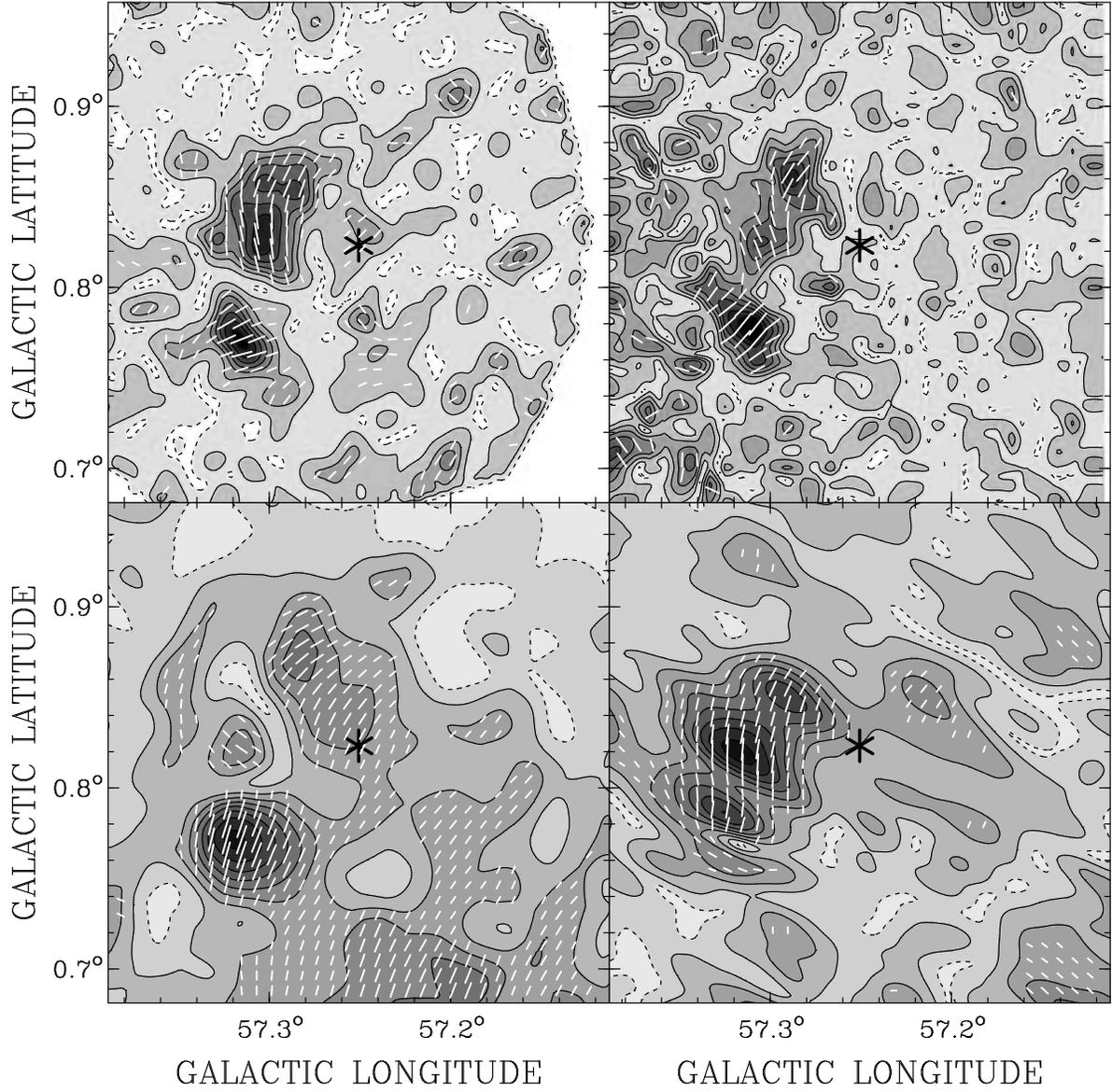}}
\caption{\label{fig:pol} Polarized intensity maps of the SNR G57.2+0.8 at 10.45~GHz (top left), 8.35~GHz
(top right), 4.85~GHz (bottom left), and 1420~MHz (bottom right). Polarization vectors in received E-field 
direction are overplotted in white. The black asterisk marks the location of the magnetar
SGR\,1935+2154.}
\end{figure}

Images of polarized intensity with overlaid vectors in E-field direction are shown in 
Figure~\ref{fig:pol}.
Only the prominent shell is significantly linearly polarized. At 10.45~GHz the 
polarization signal is divided in two parts, with a ``depolarization canal'' \citep{have04}
between them. Both parts meet with an approximate $90\degr$ separation in polarization angle.
A depression in total power is noticeable there, too. At 1420~MHz and 8350~MHz the total power 
emission structure is very similar with the same emission depression. In 
polarization at 1420~MHz, but also at 4850~MHz, the bottom
polarization feature is very similar to 10450~MHz, but the top part
is divided into two parts. The two parts are separated by another
``depolarization canal'' and at 4850~MHz both parts meet with an approximate $90\degr$ angle 
separation. At 8350~MHz the polarized intensity image is similar to 1420~MHz. 
Percentage polarization is at a few percent independent of frequency 
(see Table~\ref{tab:fluxes}), which is low for a typical shell-type SNR.
One reason might be the low angular resolution compared to the size of the SNR
so that intrinsic polarization angles average out within the observing beam and
reduce the fractional polarization.

\begin{figure}[!ht]
\centerline{
  \includegraphics[bb = 68 125 512 405,width=16cm]{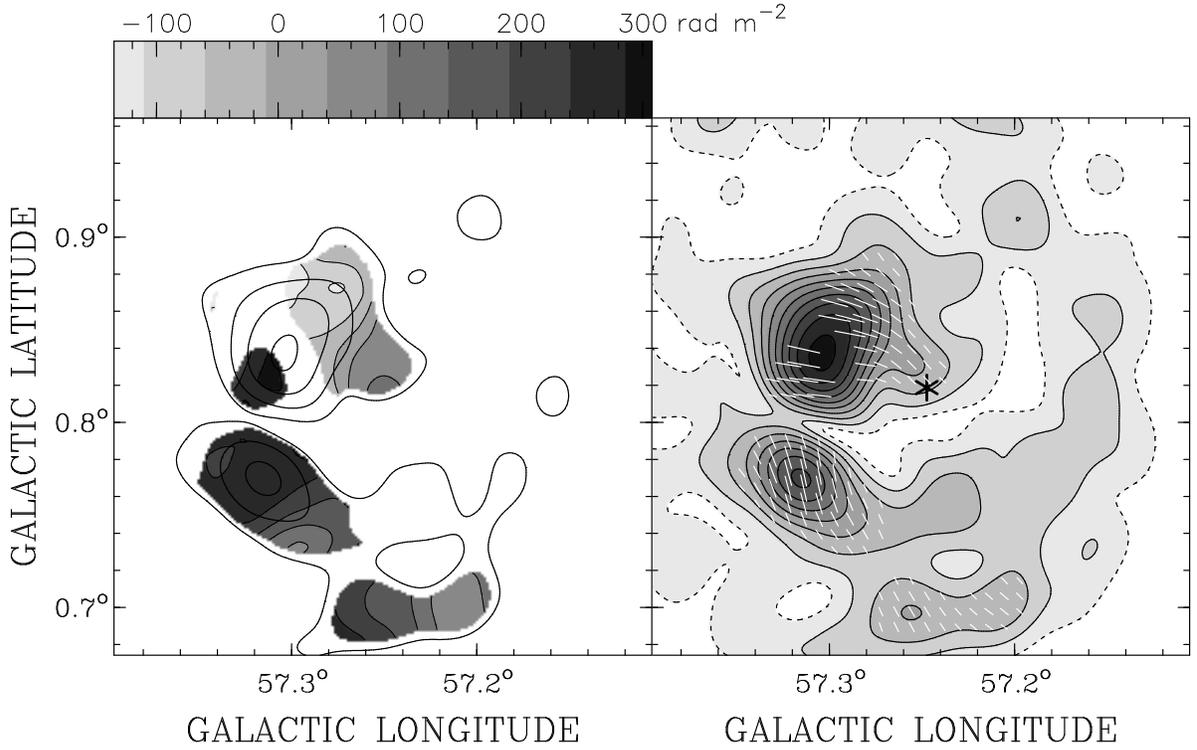}}
\caption{\label{fig:rmmap} Left: Map of rotation measure calculated between 4850~MHz and 10450~MHz at a
common resolution of $2\farcm6$. Right: Map of polarized intensity at 10450~MHz at $2\farcm6$. Overlaid are
vectors in B-field direction corrected for Faraday rotation. The black asterisk marks the location of the magnetar
SGR\,1935+2154.}
\end{figure}

We calculated a rotation measure map between 4850~MHz and 10450~GHz at the 
resolution of the 4850~MHz map displayed in Figure~\ref{fig:rmmap} . We did not
include the polarization observations 
at 8350~MHz, because of the high noise in the left part
of the image, outside the shell, and the noise gradient between the Eastern and 
Western parts of the observation. However, we used the 8350~MHz data to determine
the direction of rotation and therefore solve the ambiguity at 4850~MHz. In
the PI images in Figure~\ref{fig:pol}, in the bottom and centre part of the polarized
emission structure, the polarization angle is rotating clockwise, leading to
positive rotation measures; in the top part the polarization angle is rotating
counter-clockwise resulting in a negative RM. The rotation measure calculated between 
4850~MHz and 10.45~GHz is remarkably constant 
over the bottom PI feature and the bottom part of the top feature (see Figure~\ref{fig:rmmap})
indicating that this ``depolarization canal'' is an intrinsic
emission feature. There is a very steep RM gradient between the two parts of the top 
feature from about +240, like most of the SNR, to below $-100$~rad\,m$^{-2}$. This steep 
RM gradient causes the lack of polarized emission at 4850~MHz, but only 
causes a reduction in PI at 10450~MHz, 8350~MHz and 1420~MHz. A rotation measure of 
350~rad\,m$^{-2}$ would cause a rotation of only $17\degr$ at 10450~MHz,
$26\degr$ at 8350~MHz, and almost $80\degr$ at 4850~MHz.  
At the two higher frequencies the Faraday rotation is rather low and only causes a slight
depolarization. But at 4850~MHz the RM gradient seems to be just right 
to cause full depolarization within the beam.

It is remarkable that despite the high rotation measure of more than
$+200$~rad\,m$^{-2}$ over the bottom polarization
feature, which would rotate the polarization angle at 1420~MHz by more than 
500\degr, this feature does not seem to be ``Faraday-thick'' at this frequency.
Here, the expression Faraday-thick refers to an effect similar to a medium
being optically thick or opaque. A medium is considered to be opaque if photons
cannot pass through the medium without being absorbed. A medium is
considered to be Faraday thick if linearly polarized emission cannot pass through
it without being depolarized.

In Figure~\ref{fig:rmpuls} we display a sample RM determination for one pixel
in this area after convolving all observations to a common resolution of 
$2\farcm6$. We made four maps around 1420~MHz for the four frequency bands
provided by the DRAO Synthesis Telescope. A combined fit of all observed polarization
angles results in $RM = +233 \pm 2$~rad\,m$^{-2}$. If we use only the four bands of the
DRAO ST we get $RM = +187 \pm 54$~rad\,m$^{-2}$ and for 4850 and 10450~MHz
only, we get $RM = +226 \pm 28$~rad\,m$^{-2}$. This indicates that only smooth foreground 
Faraday rotation can explain the observed rotation measure and lack of depolarization with
wavelength. If there were strong 
internal effects, the mix of synchrotron emission and Faraday rotation would highly 
depolarize the emission at 1420~MHz, making it Faraday thick at this frequency, and a combined fit would not be possible. 
Therefore the internal magnetic field in this part of the SNR must be almost 
perpendicular to the line of sight.

The magnetic field vectors projected to the plane of the sky are shown in 
Figure~\ref{fig:rmmap} at the resolution of the 4850~MHz measurement of
$2\farcm6$. The vectors have been corrected
for Faraday rotation. If we assume that the bright radio feature represents 
the expanding shell of a supernova remnant,
those vectors are tangential for the bottom part and radial for the top part. 
This is very unusual and will be difficult to explain. In addition there is
the steep RM gradient in the radial magnetic field area. The gradient seems to 
be mostly along the magnetic field direction displayed in Fig.~\ref{fig:rmmap}.
This can be explained by the magnetic field lines gradually 
bending over along the line of sight, and therefore producing significant
internal Faraday rotation. 

\subsection{HI Data}

We searched the HI data set from the CGPS and the VGPS 
\citep{stil06} for HI shells, cavities, or filaments that might
be related to G57.2+0.8 or a possible HI absorption signal. The SNR is with a 
peak brightness of about 12~K above the background too faint to produce a 
traditional absorption profile. From a comparison of HI absorption profiles
of SNRs in Stokes I, Q, and U, \citet{koth04} found that Stokes
I profiles show excess brightness temperature fluctuation, which they attributed
to emission contribution from small clouds within the beam. They suggested
that for a reliable $3\sigma$ HI absorption signal the absorbed source requires 
a peak brightness of at least 20~K. Since the resolution of their observations 
are similar to ours and they also observed towards the Galactic plane, their 
result is immediately applicable.

There are two velocity regimes, however, where 
we find strong HI self-absorption (HISA) signals nearby, indicating cold and dense 
foreground gas. In the inner Galaxy, because of the distance ambiguity 
(see Figure~\ref{fig:hi+rc}), we 
find HISA when there is a bright background of HI emission at the far distance and a cold and 
dense foreground at the near distance for the same radial velocity \citep{gibs05}. Even 
though the peak brightness of G57.2+0.8 is quite low, only 12~K above the background,
we find HI absorption signals towards the SNR peak in the velocity regimes that
show the HISA signal (Figure~\ref{fig:abschan}). For the absorption profile in 
Fig.~\ref{fig:hi+rc} we averaged the HI signal within the beam area of the total power peak
and subtracted a background averaged over an elliptical ring surrounding the shell seen
in total power. A comparison of the HISA velocity ranges with the
Galactic rotation curve in the direction of G57.2+0.8 (Figure~\ref{fig:hi+rc}) gives a lower 
limit for the distance of about 4.5 kpc. We cannot get an upper limit from the lack of 
absorption at other velocities due to the low surface brightness of the SNR. 

In the velocity range between about $-47$ and $-55$~km\,s$^{-1}$ we found a hole, or a thick-walled
structure in the HI emission that fits the SNR, in particular the bright shell, very well 
(Figure~\ref{fig:hichan}). There does not seem to be any features in velocity that may indicate
the cap of an expanding HI shell. In Figure~\ref{fig:hichan} at the lower negative velocities
a cloud appears that turns into one thick filamentary structures that seem to wrap around
the bright shell to the left and another cloud to the north-west 
($-44.4$~km\,s$^{-1}$ $\le$ V$_{LSR}$$\le$ $-48.0$~km\,s$^{-1}$). Towards higher negative velocities
the north-west cloud disappears and the thick filament turns into a large constant surface brightness
cloud, giving the impression that the SNR is sitting in a gap of HI emission.
This also can
explain the appearance of the SNR since there seems to be more HI material to the left and an opening
to the bottom-right. To the top-left the SNR is bright indicating more material to interact with,
while to the bottom-right the emission is diffuse indicating either the expansion along a 
density gradient or into a cavity. Towards even higher negative velocity this HI gap becomes
bigger and eventually disappears.

In Figure~\ref{fig:prof} we display the radial emission profile of the SNR shell centered at the 
geometric centre of the SNR. It is compared with a radial profile of the HI emission
calculated from an HI map that combines the velocity channels between $-44.4$ and 
$-50.4$~km\,s$^{-1}$ displayed in Figure~\ref{fig:hichan}. These data have been taken from the 
VGPS \citep{stil06}, because of its better resolution. For the continuum and polarization
study the CGPS data were used because of their higher sensitivity.

The relative structures of the two radial profiles seem to show that the
SNR is located inside the HI hole indicated in the HI profile.  
This suggests that the SNR either produced this 
HI cavity by ionizing and sweeping up the material or this HI structure is an HI shell, maybe
a stellar wind bubble produced by the supernova's progenitor star.
This gives the SNR a systemic velocity in the range between $-44$ and $-51$~km\,s$^{-1}$ and
places the SNR in the far Outer Galaxy.

\section{Discussion}

\subsection{Distance to SNR G57.2+0.8}

\citet{surn16} tried to determine a distance to G57.2+0.8 via HI absorption measurements
using data from the VGPS.
However, in the VGPS the SNR has a peak brightness of about 12~K above background 
too faint to produce a reliable absorption profile. We note that \citet{surn16} did not use
the brightness peak of the SNR, but chose an area of lower surface brightness about
10~K above background. As already discussed in Section~3.3
to produce a reliable HI absorption spectrum towards G57.2+0.8, the absorbed source requires 
a peak brightness of at least 20~K. This means that a lack of 
absorption in a velocity range that displays the presence of HI emission does
not necessarily mean that this HI gas is behind the radio source. In addition,
their chosen area for the off-profile is far too small. Larger areas have to be chosen to
average out emission of smaller clouds that just happen to be at the position of the
off-profile and could give the impression of an absorption signal where none exists
in particular for low surface brightness radio sources.
This is the case for the two absorption signals that \citet{surn16} claimed to have
found in their study. We inspected the HI data from the VGPS and found a cloud at the 
off-position for those velocitites but no real absorption. Therefore their distance estimate
is clearly not supported by the data.

The HI absorption signals from G57.2+0.8 we found, which correlate very well in velocity with the 
nearby HISA features, 
indicate a distance beyond 4.5~kpc for the SNR, which is beyond the tangent point in this 
direction of our Galaxy. 

\begin{figure}[!ht]
\begin{minipage}{9cm}
 \includegraphics[bb = 580 905 1030 1335,width=9cm]{f6a.eps}
\end{minipage}
\hfill
\begin{minipage}{9cm}
  \includegraphics[bb = 70 185 535 618,width=9cm]{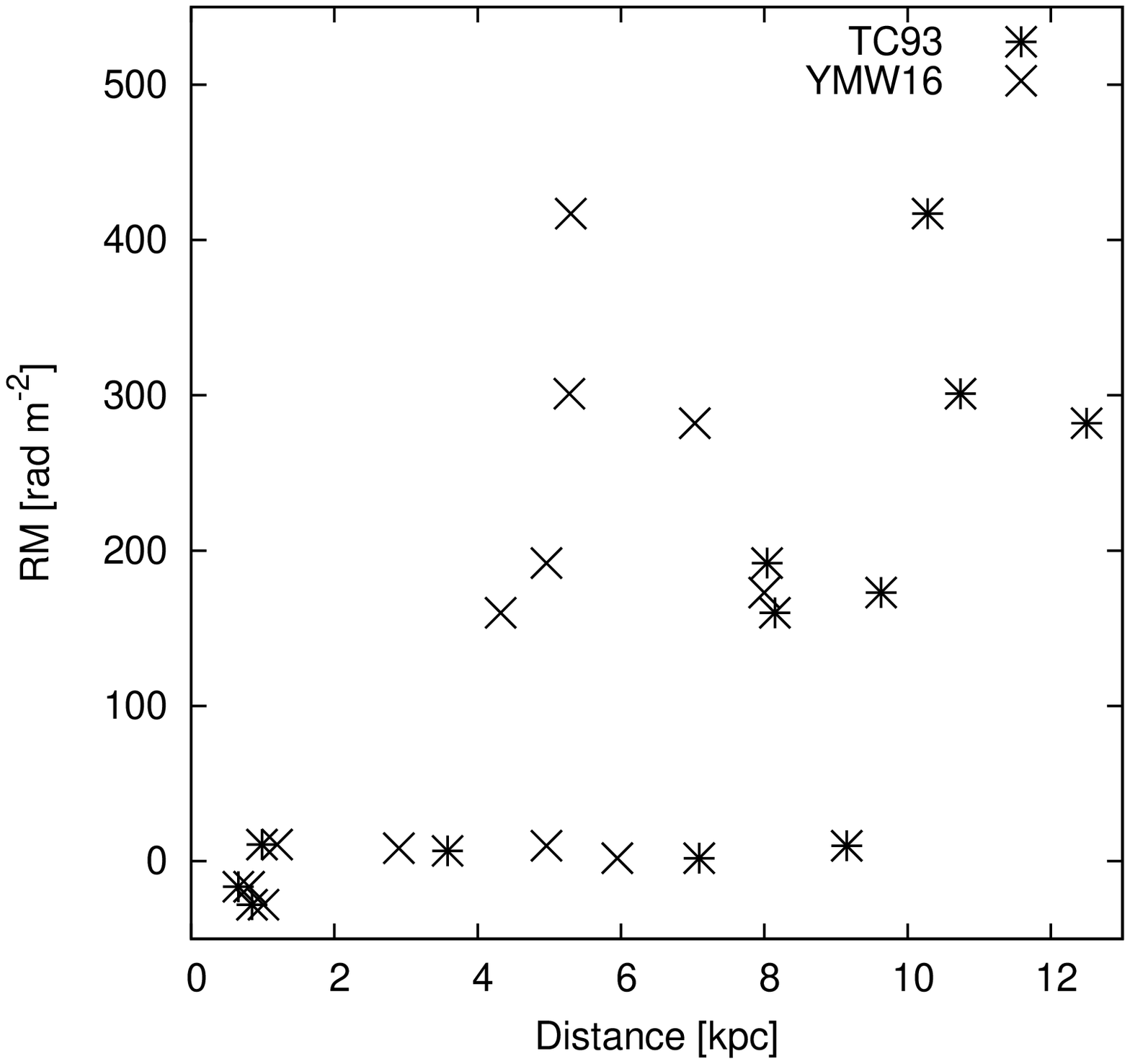}
\end{minipage}
\caption{\label{fig:rmpuls} Left: Sample RM determination for one pixel in the bottom polarization feature. The inset at 
the top left is zoomed in on the four frequency bands around
1420~MHz provided by the DRAO synthesis telescope. Right: A plot of pulsar rotation measures as a function of dispersion 
measure distance taken from the ATNF Pulsar Database 
\citep[http://www.atnf.csiro.au/research/pulsar/psrcat/, ][]{manc05}. All pulsars with known rotation measures within $5\degr$ of 
G57.2+0.8 are displayed. The rotation measures
were determined by \citet{hami87}, \citet{weis04}, \citet{lync13}, and \citet{yan11}. The dispersion measure distances were
determined with the models by \citet{tayl93} and \citet{yao17}.}%\footnote{http://www.atnf.csiro.au/research/pulsar/psrcat/}
\end{figure}

\begin{figure}[!ht]
\begin{minipage}{9cm}
  \includegraphics[bb = 65 175 540 635,width=9cm]{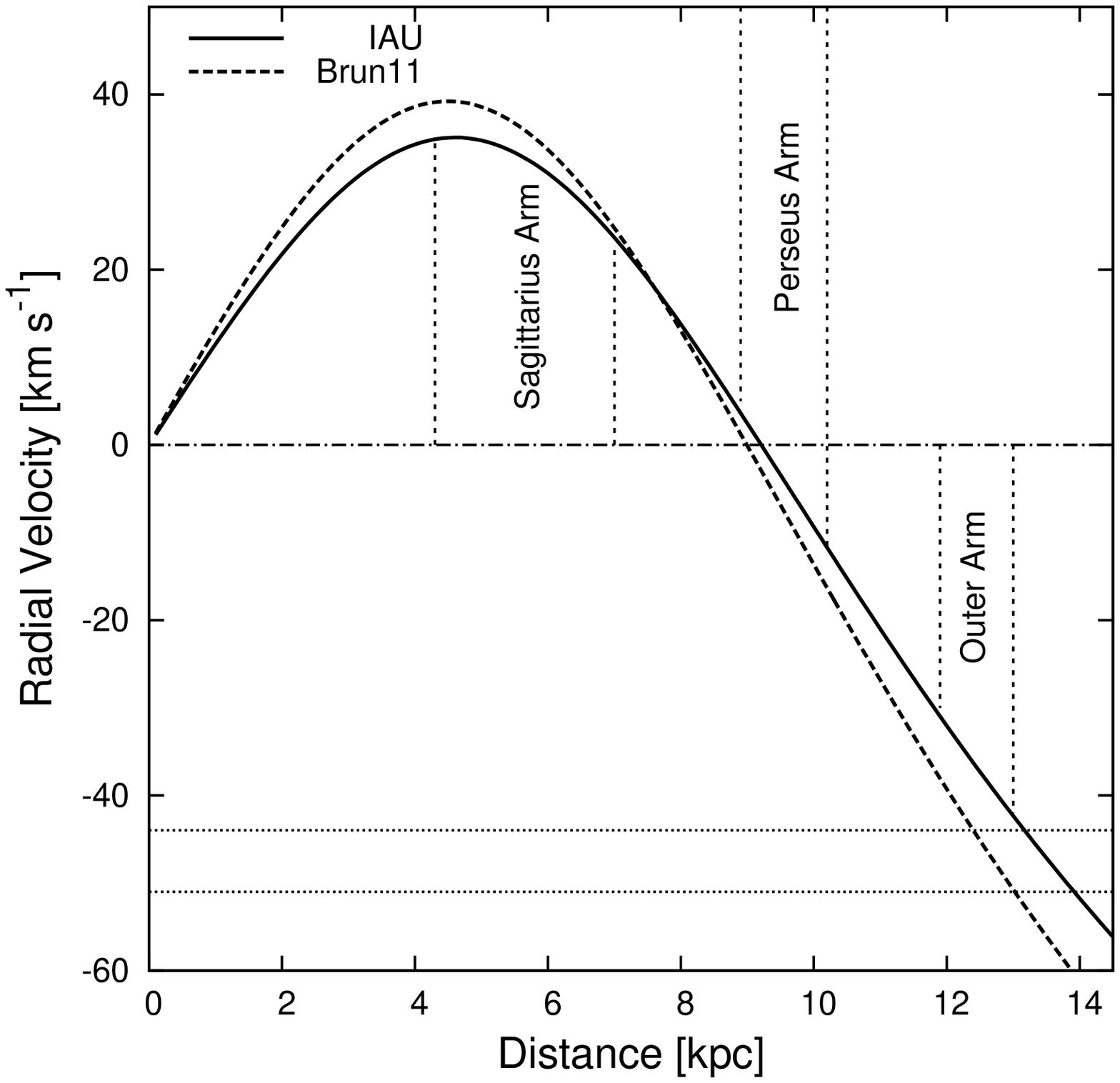}
\end{minipage}
\hfill
\begin{minipage}{9cm}
  \includegraphics[bb = 65 175 545 640,width=9cm]{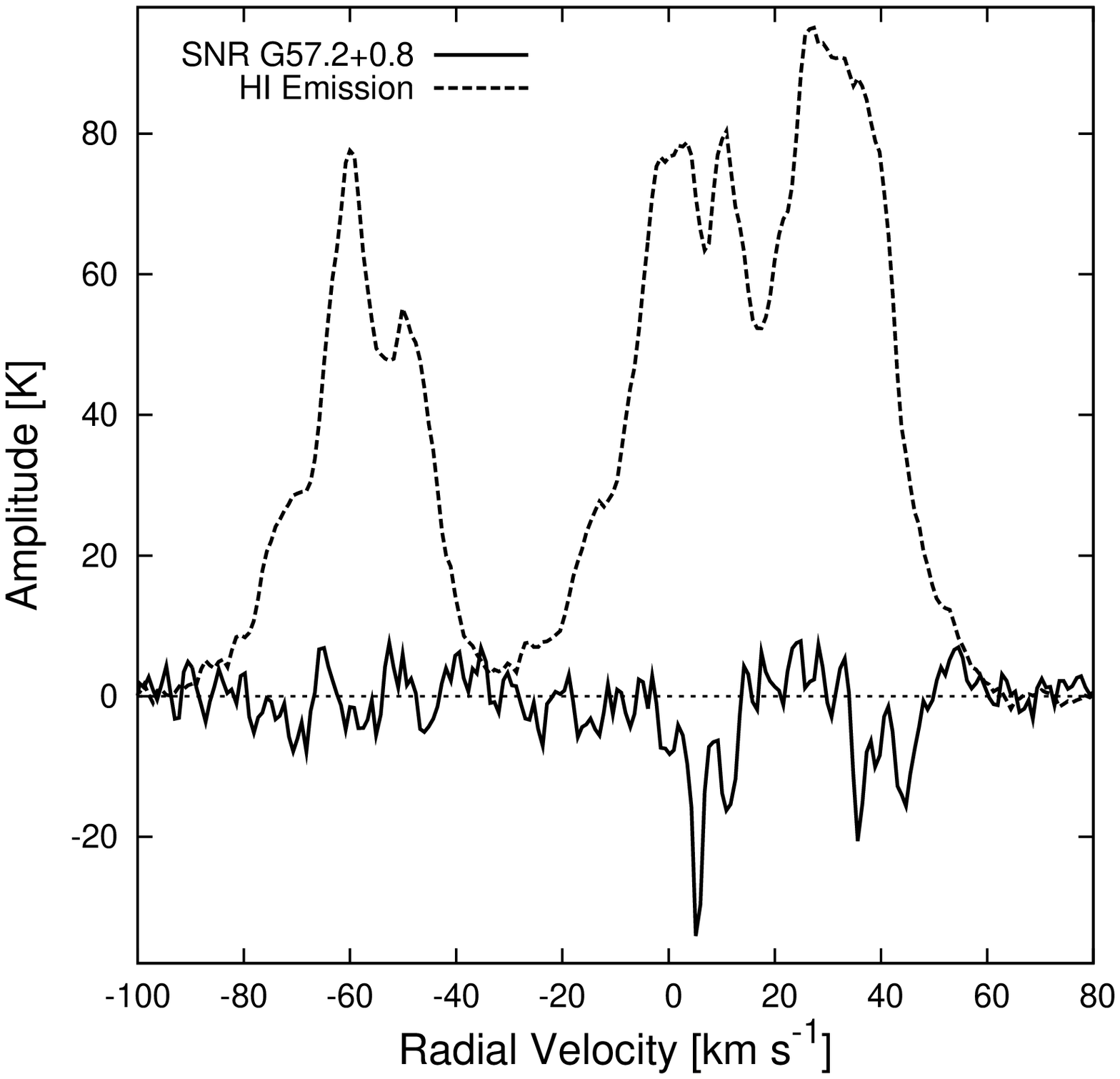}
\end{minipage}
 \caption{\label{fig:hi+rc} Left: Velocity field in
 the direction of G57.2+0.8 assuming a flat rotation curve for the Galaxy. The flat rotation model with the IAU 
 endorsed values for the sun's Galacto-centric distance of $R_\odot = 8.5$~kpc and the Sun's orbital velocity
 of $v_\odot = 220$~km\,s$^{-1}$ is plotted as a solid line and the newly determined values of 
 $R_\odot = 8.3$~kpc and 
 $v_\odot = 246$~km\,s$^{-1}$ determined by \citet{brun11} as a dashed line. Proposed locations of the 
 spiral arms 
 are indicated as is the
 proposed systemic velocity range of the SNR (see text). Right: HI absorption profile of the SNR 
 G57.2+0.8 and the HI emission profile averaged around the SNR, which was also used as the background
 emission to produce the absorption profile (see text). The absorption profile has been multiplied by a
 factor of 2 to enhance the contrast relative to the emission profile.}
\end{figure}

\begin{figure}[!ht]
\centerline{
  \includegraphics[bb = 68 68 568 360,width=18cm]{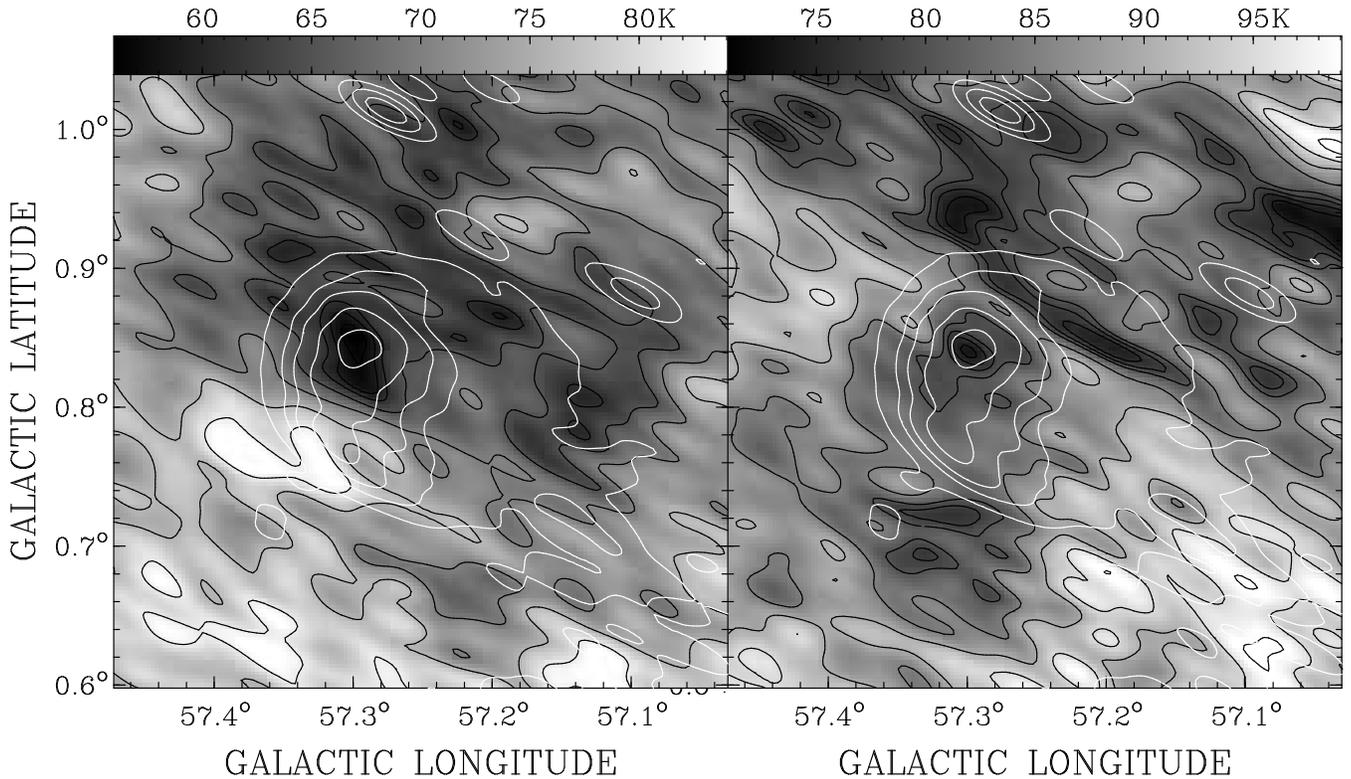}}
\caption{\label{fig:abschan} HI channel maps averaged over the absorbed velocity ranges from 
5 to 7~km\,s$^{-1}$ (left) and from 35 to 36.5~km\,s$^{-1}$ (right) taken from the CGPS
database. The total power emission at 1420~MHz is indicated by white contours.}
\end{figure}

\begin{figure}[!ht]
\centerline{
  \includegraphics[bb = 40 69 510 696,width=16cm]{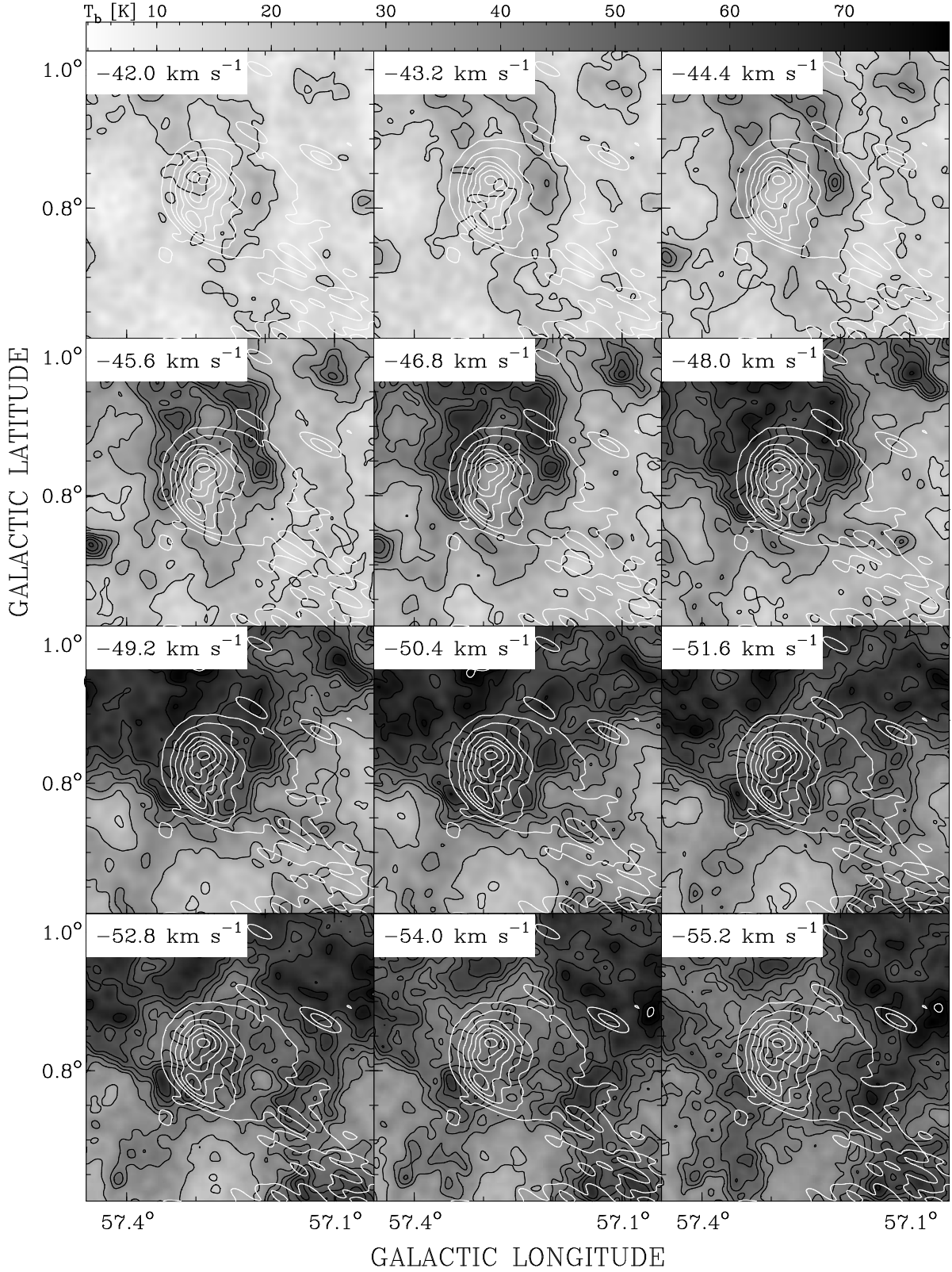}}
\caption{\label{fig:hichan} HI channelmaps in the velocity range close to the SNR's proposed
systemic velocity taken from the VGPS \citep{stil06}. White contours indicate the total power 
emission at 1420~MHz taken from the CGPS \citep{tayl03}.}
\end{figure}

\begin{figure}[!ht]
\centerline{
  \includegraphics[bb = 78 210 530 590,width=10cm]{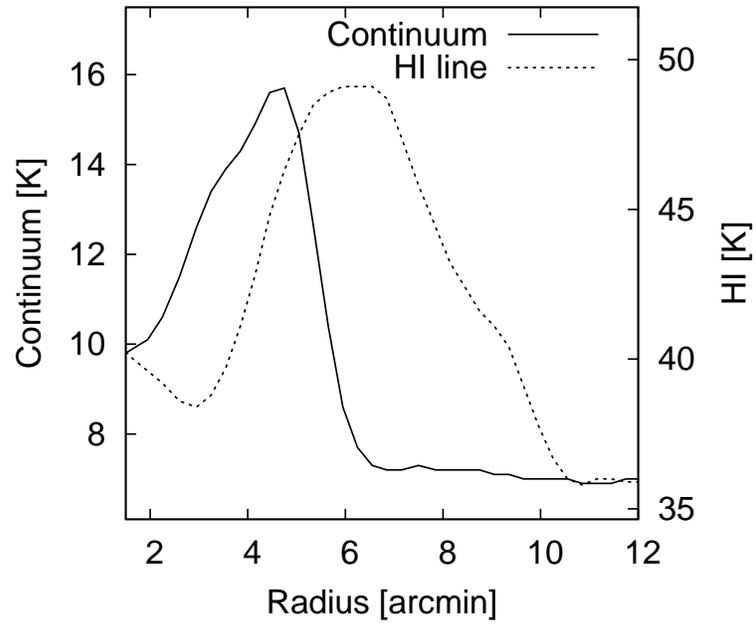}}
\caption{\label{fig:prof} Radial profiles of the radio continuum and HI emission centered at the
geometric centre of the SNR. The left Y-axis labels represent the Continuum
emission and the right Y-axis labels the HI emission at 1420~MHz. The radial profile was calculated over the 
SNR shell only.}
\end{figure}

In our polarization study we also found a very high foreground rotation measure of more than 
$+200$~rad\,m$^{-2}$. We can compare this RM with measured values for pulsars in this area
of the sky. In Figure~\ref{fig:rmpuls} we display the rotation measure of all pulsars
within $5\degr$ of G57.2+0.8 as a function of their dispersion measure (DM) distance taken from the ATNF Pulsar Catalogue 
\citep[http://www.atnf.csiro.au/research/pulsar/psrcat/, ][]{manc05}, version 1.56 (accessed March 31, 2017). 

Recently, a new Galactic electron density distribution model was introduced to this catalogue with which dispersion measure
distances to pulsars are calculated \citep[YMW16,][]{yao17}. Previously, the Taylor \& Cordes model was used 
\citep[TC93,][]{tayl93}. 
Since there is a large discrepancy between the dispersion measure distances calculated with both models in the
direction of G57.2+0.8, we
display the results for both of them in Figure~\ref{fig:rmpuls}. The major difference between them in this direction of our Galaxy
seems to be at what Galactic longitude the Sagittarius arm is contributing to the electron density along the line of 
sight. In the YMW16 model the Sagittarius arm is a dominant foreground feature in the direction of G57.2+0.8, while
it is not present in the TC93 model above about 52$\degr$ of Galactic Longitude.

In general, the foreground RM in this direction seems to be small and negative for local objects 
and highly positive for large distances (see Figure~\ref{fig:rmpuls}). All pulsars with RMs above 
$+200$~rad\,m$^{-2}$ are beyond 10~kpc using the TC93 model and beyond 5~kpc using the YMW16 model.
Therefore, we conclude that 5~kpc is a lower limit for the SNR, which is similar to our
HI absorption study. However, there is a lot of scatter in the plot, which is not surprising since 
the foreground is changing significantly towards the Galactic plane and Sagittarius arm.

The closest radio pulsar, PSR~B1930+22, is only about 40' away, at Galactic coordinates 
($\ell,b$) = (57.35$\degr$, $+1.55\degr$). It shows $RM = 173\pm 11$~rad\,m$^{-2}$ 
and a DM distance of 9.6~kpc for the TC93 model and 8.0~kpc for the YMW16 model. There is
an independent distance estimate of 10.9~kpc, based on HI absorption measurements \citep{verb12}.
A comparison of the foreground RM of the pulsar and the $+233$~rad\,m$^{-2}$ we fitted for the SNR
indicates a lower limit of 10.9~kpc for the SNR's distance. Assuming that the average foreground magnetic field
parallel to the line of sight is the same for both PSR~B1930+22 and SNR~G57.2+0.8, we can determine the SNR's
foreground dispersion measure from the DM to RM ratio of the pulsar to be DM$_{\rm SNR} \approx 290~$cm$^{-3}$\,pc.
This results in dispersion measure distances of 9~kpc and 15~kpc for the YMW16 and TC93 models, respectively.
Although all of these calculations are quite uncertain, they nevertheless indicate a very large distance well beyond the
5~kpc lower limits determined with the other methods for the SNR G57.2+0.8.

We found a systemic velocity between $-44$ and $-51$~km\,s$^{-1}$ for the HI structure, which we believe
is associated with the SNR G57.2+0.8. If we use a flat rotation curve for the Galaxy with the IAU endorsed
values for the sun's Galacto-centric distance of $R_\odot = 8.5$~kpc and the Sun's orbital velocity
of $v_\odot = 220$~km\,s$^{-1}$ this velocity interval translates to a distance between 13.2 and 14.0~kpc.
With the newest values of $R_\odot = 8.3$~kpc and $v_\odot = 246$~km\,s$^{-1}$ determined by \citet{brun11}
the distance interval would be 12.4 to 13.0~kpc. It is already clear from those two determinations that
small changes in the Galactic rotation models can cause very large differences in the distance estimate.

In the outer Galaxy in the direction of G57.2+0.8 (negative velocities), 
the HI emission profile in Figure~\ref{fig:hi+rc} shows only 
one major emission feature. There is also only one spiral arm that far out in this direction, the Outer arm. 
Since the systemic velocity of G57.2+0.8 falls within the velocity range of this emission feature, the SNR
must be a resident of the Outer arm. There has been a recent distance study of star forming regions inside
the Outer spiral arm by \citet{hach15}. They determined distances to these star forming regions with trigonometric
parallax measurements of related H$_2$O masers. They also fitted a logarithmic spiral pattern to their results
to determine distance and pitch angle of the Outer spiral arm. We indicate the location of the major spiral
arms in our Galaxy in the G57.2+0.8 velocity curve in Figure~\ref{fig:hi+rc}. Those locations are based on
Figure~7 of \citet{hach15}. From their spiral pattern fit we determine a distance of $12.5\pm 1.5$~kpc to
G57.2+0.8 by placing the SNR in the centre of the Outer arm. The error contains the width of the arm, the
uncertainty in the fit of the logarithmic spiral, and its extrapolation to the location of G57.2+0.8.

This estimate agrees within errors with the result we derived from the flat rotation model with the 
$R_\odot$ and $v_\odot$ values determined by \citet{brun11}. However, there seems to be a shift towards
higher negative velocities. This can be explained by the presence of a spiral shock in the Outer arm,
which would ``push'' objects towards the Galactic centre. That would make them appear --- from our
perspective --- at higher negative velocities than predicted from a flat rotation model. This has been shown 
for the Perseus spiral arm in the second quadrant of our Galaxy \citep{robe72}.

We will proceed by using a distance of $12.5\pm 1.5$~kpc for G57.2+0.8 and assume that this SNR
is a resident of the Outer spiral arm.

\subsection{Structure and nature of G57.2+0.8}

In our linear polarization observations, we found that part of the SNR's shell displays an intrinsic tangential
magnetic field while in the other part it is radial. The observations can be explained by a 
typical shell-type remnant with a tangential field,
part of which is overlapping with a source that contains a radial magnetic field. That would explain the
depolarization canal as the place where the radial component becomes dominant and becomes stronger in
general, since in total power there is an increase towards the area with the radial magnetic field. This
second component could be the inner arc-like feature, which we can see in the high resolution GMRT map 
(Fig.~\ref{fig:tgss}).
In our other lower resolution observations those two shells blend together, giving the impression of one
thick shell. 

If
the pulsar produced a PWN-like feature, represented by the inner arc, with a dipolar magnetic field similar 
to the PWN DA\,495 \citep{koth08b}
we could explain the large internal RM towards the magnetar and the bending over towards the shell, if we
assume that we look almost into the dipole from the top or bottom. This internal dipolar magnetic field must 
be pointing away from us, since the foreground RM is highly positive and there is the steep gradient of less
than $-300$~rad\,m$^{-2}$, which must have been produced internally. An internal RM of the
same order of magnitude was found in the PWN DA\,495 \citep{koth08b}. Extended X-ray emission related to the 
magnetar SGR\,1935+2154 was found by \citet{isra16} and interpreted as a pulsar wind nebula. The small
extent of the X-ray nebula compared to the radio nebula is typical for evolved PWNe, since the X-ray emission
represents the current energy input of the pulsar and the radio emission an integration over the entire lifetime.
This would make this PWN similar to DA\,495 \citep{koth08b}.

%The total power images (Figure~\ref{fig:tp}) seem to support this model, since there is a kind of a bump
%extending from the shell towards the location of the magnetar, which can be seen in all total power images, 
%but most prominent in the high resolution 1420~MHz and 10.45~GHz images. However, the polarized emission 
%seems to be confined to the shell and there is no obvious extension towards the magnetar in the full 
%resolution PI image
%at 10.45~GHz. On the other hand for a radial magnetic field there would be a more rapid change of
%polarization angle within the beam the closer we get to the pulsar. 

%Clearly, the available observations 
%are insufficient to solve this mystery. Higher resolution
%radio observations at various frequencies plus a deep X-ray study are required to shed more light on this.

\subsection{Characteristics of the SNR G57.2+0.8}

At a distance of $d = 12.5\pm 1.5$~kpc, the SNR's radius of about 5.5' translates to 20~pc. This is the
average distance from the geometric centre to the outer edge of the bright radio shell. 
This can be seen very well
in the radial profile (Figure~\ref{fig:prof}). Those radial profiles
determined in total power and the HI from the geometric centre match remarkably well. In HI
there is a depression of about 6 K averaged over HI velocity channels of total width $\approx 7.2$~km\,s$^{-1}$. 
It is not clear from the HI channel maps or the radial profile if this is a HI shell or
an HI hole produced by the expanding shock wave of the supernova. Although in the radial
profile it looks like an HI rim around the SNR shell, this rim would be too wide for an expanding 
stellar wind bubble and the outside of that rim marks the edge of the overall HI cloud. This
becomes more clear in the higher negative velocity channels displaid in Figure~\ref{fig:hichan}.
This would make
this SNR similar to CTB\,109, which is also located inside a HI hole,
not a stellar wind bubble, and hosts a magnetar \citep{koth02,koth12}. 
In this case, the progenitor star should have been, similar to CTB\,109, a B2 or later type star, which did
not lose much material due to a stellar wind and exploded in a type II event. B1 and earlier type stars lose most of their mass
in stellar winds, in particular in later stages of their evolution. 
%Therefore the supernova explosion 
%that created the SNR G57.2+1.8 was a core-collapse explosion of
%a star that was still very massive when it exploded.

Assuming
optically thin HI emission, the hole of 6~K translates to a minimum HI column density of
$N_{HI} = 7.8 \times 10^{19}$~cm$^{-2}$ taken away by the expanding SNR. Extrapolating
our radial profile, which was calculated only over the visible bright shell that makes up
a segment of about $90^\circ$, to the total SNR, this indicates a mass for the swept up 
material of about 1000~M$_\odot$
and a mean pre-supernova ambient density of about 1.2~cm$^{-2}$. 

The simulation of radiative SNRs by \citet{ciof88} gives the following equations for
radius $R_{\rm PDS}$ and age $t_{\rm PDS}$ at the time a SNR enters the so-called pressure-driven
snowplow or radiative phase:
\begin{equation}
   R_{\rm PDS} = 14.0 \frac{E_{51}^\frac{2}{7}}{n_0^\frac{3}{7}}~~{\rm pc}~,
\end{equation}

\begin{equation}
t_{\rm PDS} = 1.33 \times 10^4 \frac{E_{51}^\frac{3}{14}}{n_0^\frac{4}{7}}~~{\rm yr}~;
\end{equation}
here we assume solar abundances for the interstellar medium into which the SNR is expanding.
Transforming equation 3.32a from \citet{ciof88} enables us to calculate the age of the SNR
from the current radius of the shockwave $R_{\rm s}$:

\begin{equation}
t = \frac{3}{4}~t_{\rm PDS} \left(\left(\frac{R_{\rm s}}{R_{\rm PDS}}\right)^{\frac{10}{3}} + 1 \right)
\end{equation}

We have estimates of the ambient medium density and the radius of the SNR, however, the explosion energy
of a supernova can potentially range from a few times $10^{49}$ to a few times $10^{51}$~erg \citep[e.g.][]{pejc15,lyma16}.
However, at the high end, explosion energies of several times $10^{51}$~erg seem to be related to more massive stars
which lost most of their material in stellar winds and explode in type Ib/Ic events. But we did not find any evidence
of a stellar wind bubble around G57.2+0.8 and assumed a type II event for this SNR.
The radio luminosity of G57.2+0.8 between
10~MHz and 100~GHz is about $L_{radio} = 5\times 10^{33}$~erg\,s$^{-1}$, assuming a constant
spectral index of $\alpha = -0.55$ over this entire frequency range. The minimum energy
required to produce this synchrotron emission is about $E_{min} = 5\times 10^{49}$~erg,
which gives us a lower limit for the explosion energy of the supernova. However, a
supernova remnant with such a low explosion energy would merge with the interstellar medium before reaching 
such a large radius \citep[][equation 4.4b]{ciof88}. In addition, SNRs in the radiative phase of their
evolution lost already a significant amount of energy through radiation. Using equation 3.15 from
\citet{ciof88}, we simulated the evolution of the energy inside the SNR for varying explosion energies.
We found that a minimum explosion energy of $3 \times 10^{50}$~erg is necessary to retain the
$E_{min} = 5\times 10^{49}$~erg, required to produce the observed radio synchrotron emission,
up to the current SNR radius of 20~pc. We assume this is a lower limit for the SNR's explosion energy.
If we use the canonical value for the energy of a supernova explosion of 
$E_{51} = 10^{51}$~erg ($3 \times 10^{50}$~erg) we calculate $R_{\rm PDS} = 13$~pc (9.2~pc) and 
$t_{\rm PDS} = 12,000$~yr (9,200~yr).
A comparison of $R_{\rm PDS}$ with the radius of G57.2+1.8 of 20~pc clearly indicates that
this SNR already entered the radiative phase of its evolution. Using equation~3
results in a current age of about 41,000~yr (95,000~yr) for G57.2+1.8. We proceed to assume an
explosion energy of $10^{51}$~erg for the supernova explosion that produced G57.2+0.8 in the following discussion.
%with an equipartition magnetic field of $B_{eq} = 50$~$\mu$G.

\subsection{Association between SGR\,1935+2154 and G57.2+0.8}

The association between the magnetar SGR\,1935+2154 and the SNR G57.2+0.8 was first proposed by
\citet{gaen14} based on its location at the SNR's geometric centre within uncertainties. It is argued
that this is a relatively uncrowded region of the Galactic plane, which implies a high probability 
for a physical association. We now discuss whether independent distance and age estimates can 
support this association.

\subsubsection{Independent Age Estimates}

We found an age of about 41,000~yr for the SNR based on its spatial dimension and a constant ambient
number density of $n_0 = 1.2$~cm$^{-3}$. The SNR is already in its radiative phase of evolution.
The characteristic age of the magnetar is $\tau = 3600$~yr \citep{isra16}, which is an order of magnitude
shorter. The authors claim that thanks to a thorough timing analysis this characteristic age can be 
considered an upper limit to the real age. However, it is extraordinarily difficult to measure the
age of a magnetar based on its rotation characteristics \citep[e.g.][]{gao16}. The determination of the actual 
age of a pulsar requires knowledge about its braking index $n$, which in turn is determined from the pulsar's
rotation characteristics including the second period derivative $\ddot{P}$
\citep[see equations 8 and 3 in][respectively]{gao16}. This requires a constant braking index and therefore
a constant $\ddot{P}$ over the lifetime of the pulsar. However, the determination of
$\ddot{P}$ is very difficult because of its small value and large timing noise \citep{zhan13,kutu14,hask15}.
In addition, $\ddot{P}$ is strongly affected by rotational instabilities such as glitches and
outbursts, which can abruptly increase $P$ and $\dot{P}$ \citep[e.g. ][]{yuan10} and is
for magnetars consequently highly variable. 
%Therefore, the large
%discrepancy between the independently determined ages of the SNR and the magnetar 

\citet{gao16} did a study of average braking indices of magnetars 
constrained by the age of their host 
supernova remnants.
Using their method \citep[equation~10 in][]{gao16} and assuming an age of 41,000 years for both,
the SNR and the magnetar, we 
calculate a mean braking index of $n = 1.2$, similar to the value \citet{gao16} calculated for 
Swift~J1834.9--0846. \citet{gao16} suggest that braking indices higher than the canonical 3, typically 
assumed for
pure dipolar fields, are caused by a decay of the magnetic field. Braking indices in the range of 
$1 \le n \le 3$ indicate an additional source of energy loss, which is attributed to a pulsar 
wind by \citet{gao16}.

Although the study by \citet{gao16} is very speculative, it provides a possible explanation for
the descripancy between
the ages determined for the SNR and the magnetar. It also adds another piece of evidence for the presence
of a pulsar wind nebula, which we used to explain the unusual intrinsic magnetic field configuration 
derived from
the linear polarization study.

\subsubsection{Independent Distance Estimates}

We found a distance of 12.5~kpc for the SNR G57.2+0.8. \citet{isra16} suggest a distance larger than 10~kpc
based on the magnetar's radio inactivity, which can be linked to the X-ray conversion efficiency. \citet{isra16}
also favour a distance much larger than 9~kpc, because of the low outburst peak luminosity at their
assumed distance of 9~kpc. They fit spectra to extensive X-ray data of the magnetar observed with {\em Chandra}
and {\em XMM-Newton} and found typically foreground absorption column densities of $N_H$ between 1.6 and 
$2.0 \times 10^{22}$~cm$^{-2}$. The all-sky HI column density study by \citet{dick90}, which uses low
resolution HI data and assumes optically thin HI emission, results in a total column density of 
$N_H = 1.2 \times 10^{22}$~cm$^{-2}$ to the edge of the Galaxy in the direction of G57.2+0.8. Using our CGPS HI
data towards the centre of the SNR results in $N_H = 1.1 \times 10^{22}$~cm$^{-2}$. \citet{dick90}
suggest that HI column densities derived by assuming optically thin emission should be multiplied by 1.1 to 1.3
to account for HI self absorption. This still results in values lower than the foreground absorbing column
determined with X-ray observations, but certainly indicates a large distance in the Outer Galaxy. 

\citet{surn16} used the empirical relation between $N_H$ and the dispersion measure $DM$ determined by
\citet{he13} to calculate a foreground $DM$ for the magnetar from its absorbing HI column, resulting in a
foreground disperion measure of 420~pc\,cm$^{-3} \le DM \le 860$~pc\,cm$^{-3}$. A comparison
with the NE2001 model of free electrons in our Galaxy \citep{cord03} gives a distance larger than
12.5~kpc. Using the new YMW16 model \citep{yao17} gives a lower limit of 11.8~kpc. However, there is a very 
large scatter in the $N_H$-$DM$ diagram in \citet{he13}, in particular for high values of $N_H$. For an absorbing $N_H$ between 1 and $2 \times 10^{22}$~cm$^{-2}$ 
$DM$ values between 100 and more than 700~pc\,cm$^{-3}$ are found. Using the average $DM$ to $N_H$ ratio of the 5 closest
pulsars in Table~1 of \citet{he13} gives a foreground DM for our magnetar of $270^{+ 70}_{- 60}$~pc\,cm$^{-3}$, which 
is almost the same we got independently from our RM study of the SNR in section~4.1. Overall,
the independent distance estimates of the SNR and the magnetar support a physical
connection.

%Overall, given the results of our independent age and distance studies, we propose that there is a very high probability that 
%the SNR G57.2+0.8 and the magnetar SGR\,1935+2154
%were created in the same supernova explosion. 

\section{Conclusions}

We have presented a thorough analysis of radio continuum and linear polarization observations of the Galactic
SNR G57.2+0.8. 
%which hosts the recently discovered magnetar SGR\,1935+2154.
G57.2+0.8 displays a bright radio shell in its eastern part and a smooth emission plateau that disappears 
into the noise towards the west. This bright radio shell consists of two narrow arc-like features when seen
at high angular resolution. We found a typical overall spectral index of an evolved SNR of
$\alpha = -0.55 \pm 0.02$. The linear polarization study leads to a tangential magnetic field in the lower
part of the remnant and a radial field in the upper part, the latter of which could be explained by an
overlapping PWN-like feature related to the magnetar SGR\,1935+2154. The discovery of a small X-ray pulsar 
wind nebula by \citet{isra16} supports this assumption. 

We derived an Outer spiral arm location for
the SNR at a distance of 12.5~kpc. This gives the SNR a radius of 20~pc and the proposed age of about 41,000~yr
make it a highly evolved radiative SNR. 
%Based on the atomic environment, which does not show evidence of a stellar
%wind bubble we propose a B2 or later type progenitor star that exploded in a type II supernova without
%losing a significant amount of mass during its lifetime in a wind.
Based on the magnetar's location at the geometric centre of the SNR and the proximity of the
indepedently determined distances we propose a physical association between G57.2+0.8 and
SGR\,1935+2154.

\acknowledgments

We like to thank Dr. Tom Landecker and 
Dr. Ralph Eatough for carefull reading of the manuscript. We also like to express
our special thanks to Harsha Blumer and Dr. Ralph Eatough for extremely valuable 
discussions about magnetars, their braking indices, and SNR-magnetar pairs.
The Dominion Radio Astrophysical Observatory is a National Facility
operated by the National Research Council Canada. The Canadian Galactic
Plane Survey is a Canadian project with international partners, and is
supported by the Natural Sciences and Engineering Research Council of Canada
(NSERC). The Dunlap Institute is funded through an endowment established 
by the David Dunlap family and the University of Toronto. B.M.G. acknowledges 
the support of NSERC through grant RGPIN-2015-05948, and of the Canada Research Chairs program.
This research is based on observations with the 100-m telescope of the Max Planck 
Institut f\"ur Radioastronomie at Effelsberg. The National Radio Astronomy Observatory 
is a facility of the National Science Foundation operated under cooperative agreement 
by Associated Universities, Inc. This research made use of the ATNF Pulsar Database,
which can be found at http://www.atnf.csiro.au/research/pulsar/psrcat/.
We thank the staff of the GMRT that made the observations for the TGSS possible. GMRT is 
run by the National Centre for Radio Astrophysics of the Tata Institute of Fundamental 
Research.

\bibliographystyle{aa}
\bibliography{kothes}

\begin{deluxetable}{rccc}
   \tablewidth{0pc}
   \tablecolumns{4}
\tablecaption{\label{tab:fluxes} Integrated total flux densities $S$ and
polarized intensities $PI$ of G57.2+0.8 at frequencies $\nu$. The 74~MHz
flux density was integrated from a map taken from the VLSSr \citep{vlss}.}
\tablehead{$\nu$ [MHz] & $S$ [Jy] & $PI$ [mJy] & $\%$-pol.}
\startdata
74 & 6.50$\pm$1.70 & --- &--- \\
408 & 2.60$\pm$0.40 & --- &--- \\
1420 & 1.37$\pm$0.08 & 52$\pm$6 &  4$\pm$1 \\
2639 & 0.85$\pm$0.08 & 15$\pm$8 & 2$\pm$1 \\
4850  & 0.68$\pm$0.07 &  22$\pm$6 & 3$\pm$1 \\
8350 & 0.54$\pm$0.05 & 11$\pm$6 & 2$\pm$1 \\
10450 & 0.50$\pm$0.05 &  30$\pm$4 & 6$\pm$1 \\
\enddata
\end{deluxetable}

\end{document}